\documentclass[sigconf]{acmart}
\usepackage{arydshln}
\usepackage{dsfont}
\usepackage{float}
\usepackage{multirow}
\usepackage{subcaption}
\usepackage[normalem]{ulem}

\AtBeginDocument{%
  \providecommand\BibTeX{{%
    \normalfont B\kern-0.5em{\scshape i\kern-0.25em b}\kern-0.8em\TeX}}}

\copyrightyear{2022} 
\acmYear{2022} 
\setcopyright{rightsretained}
\acmConference[SIGIR '22]{Proceedings of the 45th International ACM SIGIR Conference on Research and Development in Information Retrieval}{July 11--15, 2022}{Madrid, Spain.}
\acmBooktitle{Proceedings of the 45th Int'l ACM SIGIR Conference on Research and Development in Information Retrieval (SIGIR '22), July 11--15, 2022, Madrid, Spain}
\acmDOI{10.1145/3477495.3532017}
\acmISBN{978-1-4503-8732-3/22/07}

\usepackage{etoolbox}
\makeatletter
\patchcmd{\maketitle}{\@copyrightpermission}{
   \begin{minipage}{0.3\columnwidth}
     \href{http://creativecommons.org/licenses/by/4.0/}{\includegraphics[width=0.90\textwidth]{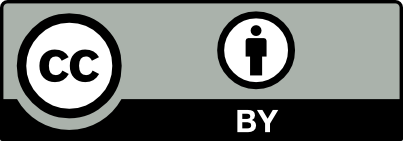}}
   \end{minipage}\hfill
   \begin{minipage}{0.7\columnwidth}
     \href{http://creativecommons.org/licenses/by/4.0/}{This work is licensed under a Creative Commons Attribution International 4.0 License.}
   \end{minipage}
  
   \vspace{5pt}
}{}{}
\makeatother

\begin{document}
\fancyhead{}

\title{LoL: A Comparative Regularization Loss over Query Reformulation Losses for Pseudo-Relevance Feedback}

\author{Yunchang Zhu}
\affiliation{%
  \institution{Data Intelligence System Research Center, Institute of Computing Technology, CAS}
  \institution{University of Chinese Academy of Sciences}
  \city{Beijing}
  \country{China}
}
\email{zhuyunchang17s@ict.ac.cn}

\author{Liang Pang}
\authornote{Corresponding author}
\affiliation{%
  \institution{Data Intelligence System Research Center, Institute of Computing Technology, CAS}
  \city{Beijing}
  \country{China}
}
\email{pangliang@ict.ac.cn}

\author{Yanyan Lan}
\affiliation{%
  \institution{Institute for AI Industry Research, Tsinghua University}
  \city{Beijing}
  \country{China}
}
\email{lanyanyan@tsinghua.edu.cn}

\author{Huawei Shen}
\affiliation{%
  \institution{Data Intelligence System Research Center, Institute of Computing Technology, CAS}
  \institution{University of Chinese Academy of Sciences}
  \city{Beijing}
  \country{China}
}
\email{shenhuawei@ict.ac.cn}

\author{Xueqi Cheng}
\affiliation{%
  \institution{CAS Key Lab of Network Data Science and Technology, Institute of Computing Technology, CAS}
  \institution{University of Chinese Academy of Sciences}
  \city{Beijing}
  \country{China}
}
\email{cxq@ict.ac.cn}

\renewcommand{\shortauthors}{Zhu and Pang, et al.}

\begin{abstract}
Pseudo-relevance feedback (PRF) has proven to be an effective query reformulation technique to improve retrieval accuracy. It aims to alleviate the mismatch of linguistic expressions between a query and its potential relevant documents.
Existing PRF methods independently treat revised queries originating from the same query but using different numbers of feedback documents, resulting in severe query drift.
Without comparing the effects of two different revisions from the same query, a PRF model may incorrectly focus on the additional irrelevant information increased in the more feedback, and thus reformulate a query that is less effective than the revision using the less feedback.
Ideally, if a PRF model can distinguish between irrelevant and relevant information in the feedback, the more feedback documents there are, the better the revised query will be.
To bridge this gap, we propose the Loss-over-Loss (LoL) framework to compare the reformulation losses between different revisions of the same query during training.
Concretely, we revise an original query multiple times in parallel using different amounts of feedback and compute their reformulation losses.
Then, we introduce an additional regularization loss on these reformulation losses to penalize revisions that use more feedback but gain larger losses.
With such comparative regularization, the PRF model is expected to learn to suppress the extra increased irrelevant information by comparing the effects of different revised queries.
Further, we present a differentiable query reformulation method to implement this framework. 
This method revises queries in the vector space and directly optimizes the retrieval performance of query vectors, applicable for both sparse and dense retrieval models.
Empirical evaluation demonstrates the effectiveness and robustness of our method for two typical sparse and dense retrieval models. 
\end{abstract}



\maketitle


\section{Introduction}

\begin{figure}
  \centering
  \includegraphics[width=\linewidth]{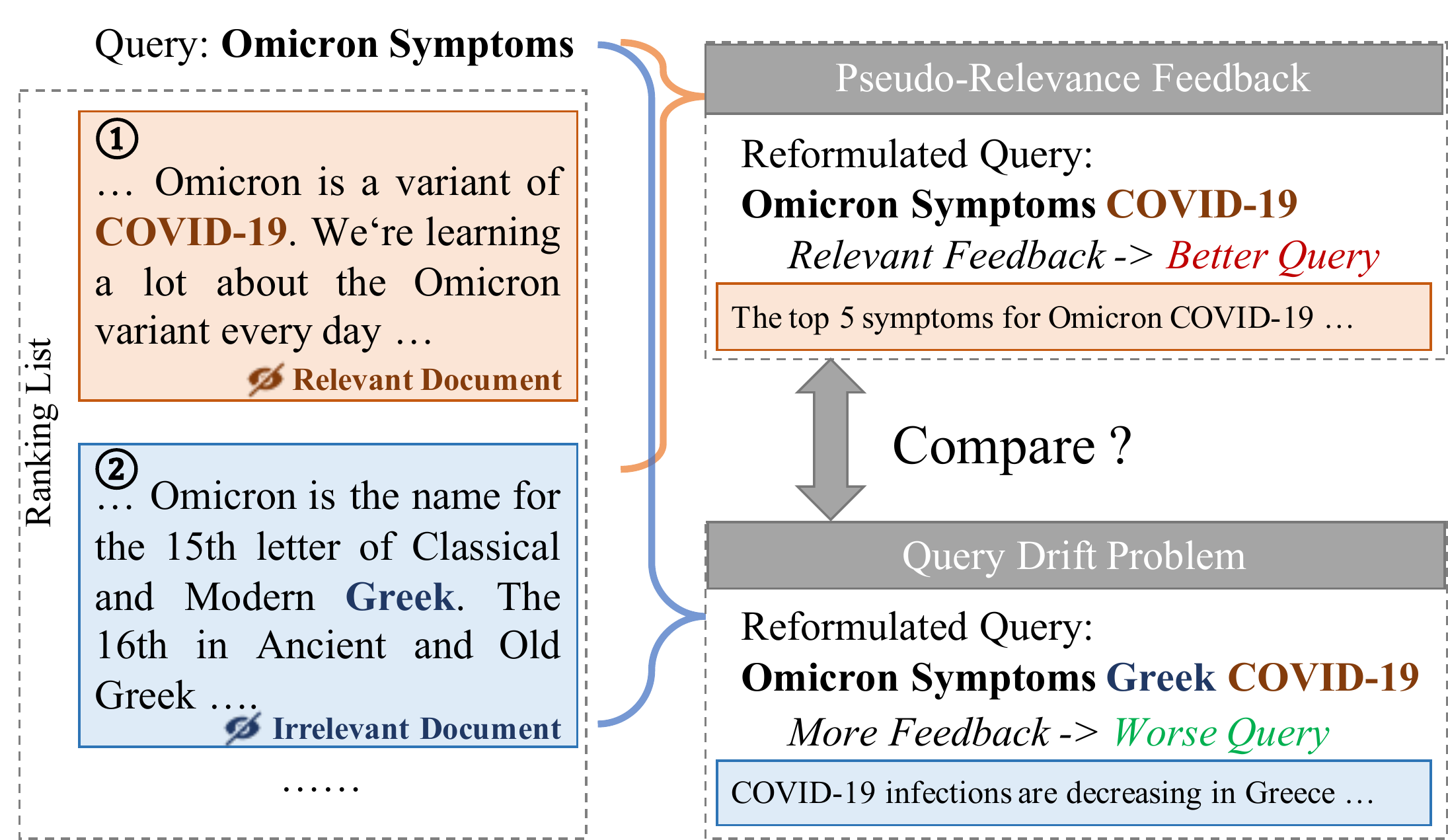}
  \caption{An example of pseudo-relevant feedback. When the top 1 document (potentially relevant) is used as the feedback, the reformulated query is better than the original. But if more documents,  such as the second (potentially irrelevant), are added to the feedback set, it may cause query drift.}
  \label{fig:example}
\end{figure}

In information retrieval (IR), users often formulate short and ambiguous queries due to the reluctance and the difficulty in expressing their information needs precisely in words~\cite{carpineto2012}.
This may specifically arise from several reasons, such as the use of inconsistent terminology, commonly known as vocabulary mismatch~\cite{furnas1987}, or the lack of knowledge in the area in which information is sought.  
Decades of IR research demonstrate that such casual queries prevent a search engine from correctly and completely satisfying users’ information needs~\cite{croft2010}.
To mitigate the mismatch of expressions between a query and its potential relevant documents, many query reformulation approaches leveraging on external resources (such as thesaurus and relevance feedback) have been proposed to revise a better query -- one that ranks relevant documents higher.

Pseudo-Relevance Feedback (PRF)~\cite{attar1977} has shown to be one of effective query reformulation technique in various information retrieval settings~\cite{manning2008, zhu2021aiso, fan2022}.
As the name implies, no real relevant feedback from users is required in PRF, which makes it more convenient and and widely studied.
In the first-pass retrieval of PRF, a small set of top-retrieved documents for an original query, called the feedback set, is assumed to contain relevant information~\cite{attar1977, croft1979}.
The ``pseudo'' feedback set is then exploited as external resources in the query reformulation process to form a query revision, which is then run to retrieve the final list of documents presented to the user.
An example is shown in Figure~\ref{fig:example} where the first document introduces the synonymous term `COVID-19' into the original query to clarify the original query that contains ambiguous `Omicron'.
Early PRF was widely studied for sparse retrieval like vector space models~\cite{rocchio1971}, probabilistic models~\cite{robertson1976}, and language modeling methods~\cite{lafferty2001, lavrenko2001rm, zhai2001, jaleel2004rm3, tao2006rmm, lv2014medmm}.
Recently, some work has shifted to apply PRF in dense retrieval of single-representation~\cite{yu2021anceprf, li2021b, li2021c} and multi-representation~\cite{wang2021a}.

Although PRF methods are generally accepted to improve average retrieval effectiveness~\cite{carpineto2012, lv2009, clinchant2013, yu2021anceprf}, their performance is sometimes inferior to the original query~\cite{mitra1998, tao2006rmm, zighelnic2008}. 
One of the causes for this robustness problem is query drift: the topic of the revision drifts away from the original intent~\cite{mitra1998}. 
This is not surprising, considering that many top-ranked documents can be irrelevant and misleading, and relevant documents may contain irrelevant information.
For example in Figure~\ref{fig:example}, adding more feedback documents, e.g. the second document, leads to a worse reformulated query, because an irrelevant document introduces the noise term `Greek' into the reformulated query which totally changes the meaning of the original query.
Therefore, PRF models need to learn to suppress the irrelevant information in the feedback set and make the most of the relevant information.
Imagine an ideal PRF model, given a larger feedback set in which both relevant and irrelevant information increase, the model should form a better query revision.

Previous studies cope with query drift mainly by adding pre-processing or post-processing~\cite{mitra1998, amati2004, lv2009a, collins-thompson2009} or leveraging state-of-the-art pre-trained language models~\cite{wang2021a, yu2021anceprf}.
However, additional processing brings more computational cost, and pre-trained language models may not necessarily learn to suppress irrelevant information for retrieval without particular supervision~\cite{ma2021}. 
Moreover, existing PRF methods optimize different revisions of the same query independently by minimizing their own reformulation losses, ignoring the {\bfseries comparison principle} between these revisions: {\bfseries the more feedback, the better the revision}, a necessary condition for an ideal PRF model.

Thus, to explicitly pursue this principle during training, we propose a conceptual framework, namely \textbf{L}oss-\textbf{o}ver-\textbf{L}oss (LoL).
This is a general optimization framework applicable to any supervised PRF method.
First, to enable performance comparisons across revisions, the original query is revised multiple times in a batch using feedback sets of different sizes.
Then, we impose a comparative regularization loss on all reformulation losses derived from the same original query to penalize those revisions that use more feedback but obtain larger losses.
Specifically, the comparative regularization is a pairwise ranking loss of these reformulation losses, where the ascending order of reformulation losses is expected to coincide with the descending order of the sizes of the feedback sets they use.
With such comparative regularization, we expect the PRF model to learn to suppress those extra increased irrelevant in more feedback by comparing the effects of different revisions.
Furthermore, we present a differentiable PRF method as a simple implementation of LoL. 
The method revises queries in the vector space, thus avoiding the hassle of natural language generation and gradient back-propagation, which makes it applicable for sparse retrieval as well as dense retrieval.
Besides, this method uses a ranking loss as the query reformulation loss, which ensures the consistency of PRF with its ultimate goal, i.e., improving retrieval effectiveness.

To verify the effectiveness of our method, we evaluate two implemented LoL models, one for sparse retrieval and the other for dense retrieval, on multiple benchmarks based on MS MARCO passage collection.
Experimental results show that the retrieval performance of LoL models is significantly better than their base models and other PRF models.
Furthermore, we prove the critical role of comparative regularization through ablation studies and visualization of learning curves.
Moreover, our analysis demonstrates that LoL is more robust to the number of feedback documents compared to PRF baselines and is not sensitive to the training hyper-parameters.

The main contributions can be summarized as follows:
\begin{itemize}
    \item A comparison principle is pointed out: the more feedback documents, the better the effect of the reformulated query. Ignoring this principle during training may cause PRF models to be misled by irrelevant information that appears in more feedback, leading to query drift.
    \item A comparative regularization is proposed to enhance the irrelevant information suppression ability of PRF models, applicable for both sparse and dense retrieval.
    \item With the help of comparative regularization, our PRF models outperform their base retrieval models and state-of-the-art PRF baselines on multiple benchmarks based on MS MARCO. We release the code at \url{https://github.com/zycdev/LoL}.
\end{itemize}

\section{Related Work}

Query reformulation is the process of refining a query to improve ranking effectiveness. 
According to the external resources used in the reformulation process, there are two categories of methods: global and local~\cite{xu1996}. 
The first category of methods typically expands the query based on global resources, such as WordNet~\cite{gong2005}, thesauri~\cite{shiri2006}, Wikipedia~\cite{aggarwal2012}, Freebase~\cite{xiong2015}, and Word2Vec~\cite{diaz2016}.
While the second category, the so-called relevance feedback~\cite{rocchio1971}, is usually more popular.
It leverages local relevance feedback for the original query to reformulate a query revision.
Relevance feedback information can be obtained through explicit feedback (e.g., document relevance judgments~\cite{rocchio1971}), implicit feedback (e.g., click-through data~\cite{joachims2002}), or pseudo-relevance feedback (assuming the top-retrieved documents contain information relevant to the user's information need~\cite{attar1977, croft1979}).
Of these, pseudo-relevance feedback is the most common, since no user intervention is required.
Although pseudo-relevance feedback (PRF) is also used for re-ranking~\cite{li2018nprf, zheng2020bertqe}, we next focus on PRF methods in sparse and dense retrieval.
Finally, we review existing efforts to mitigate query drift.

\subsection{PRF for Sparse Retrieval}

Pseudo-relevance feedback methods for sparse retrieval have a long history, dating back to Rocchio~\cite{rocchio1971}.
The Rocchio algorithm is originally a relevance feedback method proposed for vector space models~\cite{salton1975} but is also applicable to PRF.
It constructs the refined query representation through the linear combination of the sparse vectors of the query and feedback documents.
After that, many other heuristics were proposed. 
For probabilistic models~\cite{maron1960}, the feedback documents are naturally treated as examples of relevant documents to improve the estimation of model parameters~\cite{robertson1976}. 
Whereas for language modeling approaches~\cite{ponte1998}, PRF can be implemented by exploiting the feedback set to estimate a query language model~\cite{zhai2001, lv2014medmm} or relevance model~\cite{lavrenko2001rm, jaleel2004rm3}.
Overall, these methods expand new terms to the original query or/and reweight query terms by exploiting statistical information on the feedback set and the whole collection.
Besides, some recent methods expand the query using static~\cite{zamani2016} or contextualized embeddings~\cite{naseri2021ceqe}.
For instance, CEQE~\cite{naseri2021ceqe} uses BERT~\cite{devlin2019bert} to compute contextual representations of terms in the query and feedback documents and then selects those closest to query embeddings as extension terms according to some measure.
But these methods are still heuristic because they make strong assumptions to estimate the feedback weight for each term.
For example, the divergence minimization model~\cite{zhai2001} assumes that a term with high frequency in the feedback documents and low frequency in the collection should be assigned a high feedback weight. 
However, these assumptions are not necessarily in line with the ultimate objective of PRF, i.e., improving retrieval performance.

Due to the discrete nature of natural language, the reformulated query text is non-differentiable, making it difficult for supervised learning to optimize retrieval performance directly.
Therefore, \cite{montazeralghaem2020} proposes a general reinforcement learning
framework to directly optimize retrieval metrics.
To escape the expensive and unstable training of reinforcement learning, a few supervised methods~\cite{cao2008, qi2019golden} are optimized to generate oracle query revisions by selecting good terms or spans from the feedback documents. 
However, these ``oracle'' query revisions are constructed heuristically and do not necessarily achieve optimal ranking results.
Unlike all the above methods, our introduced method for sparse retrieval refines the query in the vector space, enabling direct optimization of retrieval performance in an end-to-end fashion.

\subsection{PRF for Dense Retrieval}

Dense retrieval has made great progress in recent years. 
Since dense retrievers~\cite{karpukhin2020dpr, xiong2020ance, khattab2020colbert} use embedding vectors to represent queries and documents, a few methods~\cite{wang2021a, yu2021anceprf, li2021b, li2021c} have been studied to integrate pseudo-relevance information into reformulated query vectors. 
ColBERT-PRF~\cite{wang2021a} first verified the effectiveness of PRF in multi-representation dense retrieval~\cite{khattab2020colbert}. 
Specifically, it expands multi-vector query representations with representative feedback embeddings extracted by KMeans clustering.
\cite{li2021b} investigated two simple methods, Average and Rocchio~\cite{rocchio1971}, to utilize feedback documents in single-representation dense retrievers (e.g., ANCE~\cite{xiong2020ance}) without introducing new neural models or further training.
Instead of refining the query vector heuristically, ANCE-PRF~\cite{yu2021anceprf} uses RoBERTa~\cite{liu2019roberta} to consume the original query and the top-retrieved documents from ANCE~\cite{xiong2020ance}.
Keeping the document index unchanged, ANCE-PRF is trained end-to-end with relevance labels and learns to optimize the query vector by exploiting the relevant information in the feedback documents. 
Further, \cite{li2021c} investigate the generalisability of the strategy underlying ANCE-PRF~\cite{yu2021anceprf} to other dense retrievers~\cite{lin2020tct, hofstatter2021tas}.
Although our presented PRF method for dense retrieval uses the same strategy, all the above methods are optimized to perform well on average, ignoring the performance comparison between different versions of a query.

\subsection{Coping with Query Drift}
Query drift is a long-standing problem in PRF, where the topic of the reformulated query drifts in an unintended direction mainly due to the introduced irrelevant information from the feedback set~\cite{croft1979, mitra1998}.
There have been many studies on coping with query drift. 
The strategies they typically employ include:
(\romannumeral1) selectively activating query reformulation to avoid performance damage to some queries~\cite{amati2004, cronen-townsend2004}; 
(\romannumeral2) refining the feedback set to increase the proportion of relevant information in it~\cite{mitra1998, zheng2020bertqe};
(\romannumeral3) varying the impact of the original query and feedback documents to highlight query-relevant information~\cite{tao2006rmm, lv2009a};
(\romannumeral4) post-processing the reformulated query to eliminate risky expansions~\cite{collins-thompson2009}; 
(\romannumeral5) model ensemble to fuse results from multiple models~\cite{collins-thompson2007, zighelnic2008}; 
(\romannumeral6) leveraging an advanced pre-trained language model~\cite{devlin2019bert, liu2019roberta} with the expectation that the model itself to be immune to noise~\cite{wang2021a, yu2021anceprf};
(\romannumeral7) introducing a regularization term in the optimization objective to constrain some principles~\cite{tao2006rmm, lv2014medmm}.
Our presented method belongs to the last two strategies, introducing no additional processing during inference. 
Unlike the other approaches in strategy (\romannumeral6) that count on the model to naturally learn to distinguish irrelevant information when learning query reformulation, LoL also provides additional supervision on comparing the effects of revisions.
Moreover, to the best of our knowledge, we are the first to impose comparative regularization on multiple revisions of the same query.

\section{Methodlogy}

This section describes our proposed framework for pseudo-relevant feedback (PRF) and its implementation method.
We first formalize the process of PRF and introduce its traditional optimization paradigm.
Then, we propose a general conceptual framework for PRF, namely Loss-over-Loss (LoL).
Finally, we present an end-to-end query reformulation method based on vector space as an instance of this framework and introduce its special handling for sparse and dense retrieval.

\subsection{Preliminary}

Given an original textual query $q$ and a document collection $C$, a retrieval model returns a ranked list of documents $D = (d_1, d_2, \cdots, d_{|D|})$. 
Let $F^{(k)} = D_{\le k}$ denote the feedback set containing the first $k$ documents, where $k \ge 0$ is often referred to as the PRF depth.
The goal of pseudo-relevant feedback is to reformulate the original query $q$ into a new representation $q^{(k)}$ using the query-relevant information in the feedback set $F^{(k)}$,
\begin{equation}
  q^{(k)} = \mathrm{QR}(q, F^{(k)}), \label{eq:qr}
\end{equation}
where $\mathrm{QR}$ is a query reformulation model based on PRF.

Denoting the reformulation loss of the revision $q^{(k)}$ as $\mathcal{L}_{\mathrm{rf}}(q^{(k)})$, the general form of $\mathrm{QR}$ is to optimize by minimizing the following loss function, which take multiple depths of PRF into consideration:
\begin{equation}
  \mathcal{L}(q) = \frac{1}{|K|}\sum_{k \in K}\mathcal{L}_{\mathrm{rf}}(q^{(k)}),
\end{equation}
where $K$ is the set of PRF depths that $\mathrm{QR}$ needs to learn in one epoch. 
For example, $K=\{1, 3, 5\}$ means that the loss considers the top-1, top-3, and top-5 documents in the ranked list as the feedback set, respectively.

However, $|K|$ is always set to 1 in many existing methods~\cite{yu2021anceprf}.
Specifically, existing PRF models are trained separately at each PRF depth, where $K=\{k\}$ is a constant set and
\begin{equation}
  \mathcal{L}(q) = \mathcal{L}_{\mathrm{rf}}(q^{(k)}).\label{eq:old}
\end{equation}

Taking it a little further, let $A \supseteq K$ be the set of all PRF depths that a PRF model is designed to handle, e.g., $A = \{1, 2, 3, 4, 5\}$.
If the PRF model needs to be trained jointly at all PRF depths, we can sample a random subset from $A$ as $K$ in each epoch.

\subsection{Loss-over-loss Framework for PRF}

\begin{figure}[t]
  \centering
  \includegraphics[width=0.95\linewidth]{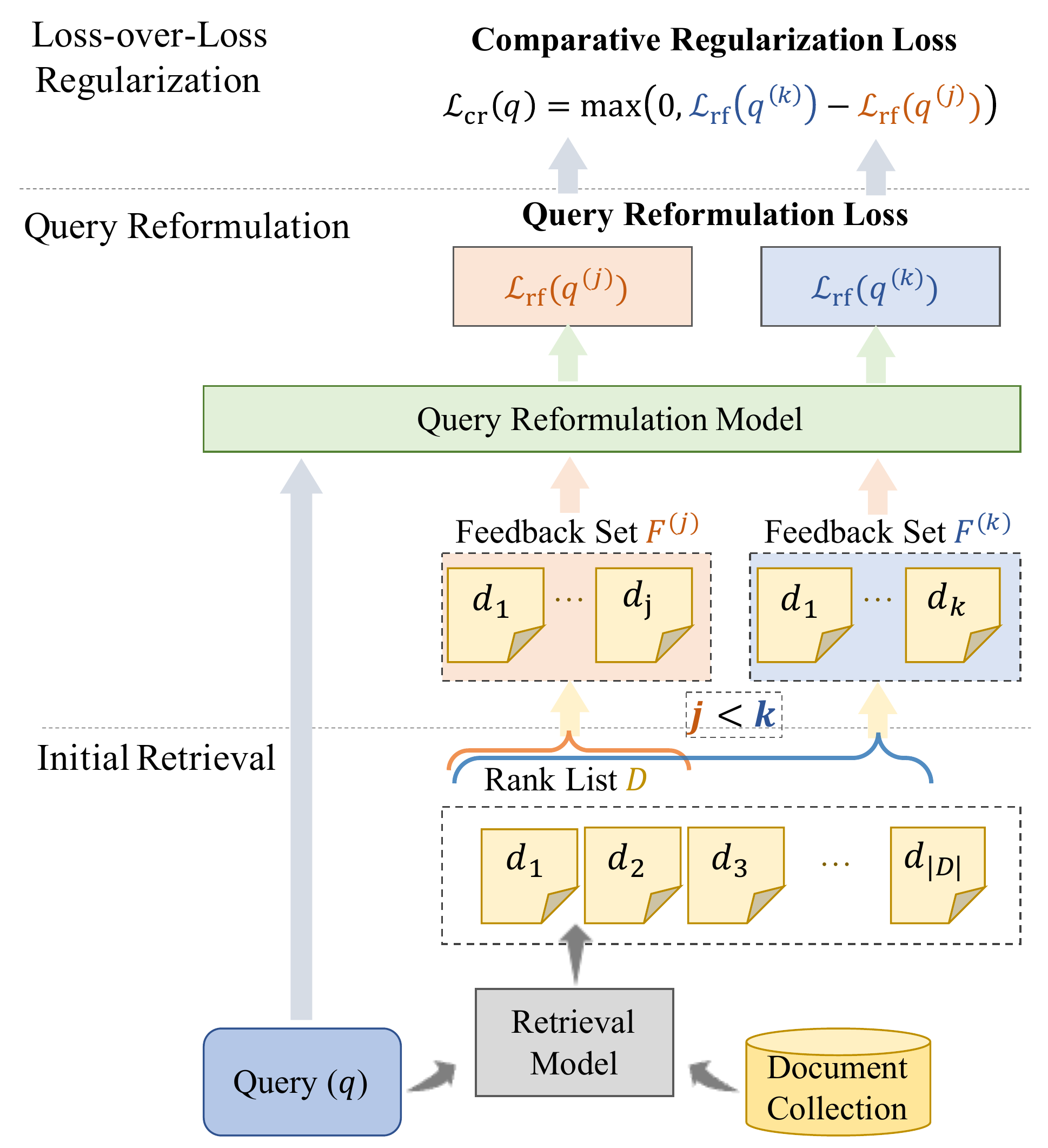}
  \caption{The overview of the LoL framework. In the initial stage, the retrieval model first generates a ranking list based on the original query. In the reformulate stage, a list of top-k documents is selected to reformulate the query. In the loss-over-loss regularization stage, a constrain is constructed to ensure more feedback documents lead to small query reformulation loss.}
  \label{fig:lol}
\end{figure}

To prevent query drift caused by the increase of (irrelevant) feedback information, we propose the Loss-over-Loss (LoL) framework. 
We first discover a comparison principle that an ideal PRF model should guarantee but was neglected in previous work. 
This principle describes the ideal comparative relationship between revisions derived from the same query but using different amounts of feedback. 
Therefore, we regularize the reformulation losses of these revisions.

Suppose $\mathrm{RI}(q, F)$ denotes the information relevant to the query $q$ in the feedback set $F$, while $\mathrm{NRI}(q, F)$ represents the information irrelevant to $q$ in $F$. 
Normally, as the PRF depth $k$ increases, both relevant and irrelevant information increase, i.e., $\mathrm{RI}(q, F^{(k+1)}) \supseteq \mathrm{RI}(q, F^{(k)})$ and $\mathrm{NRI}(q, F^{(k+1)}) \supseteq \mathrm{NRI}(q, F^{(k)})$. 
In this way, an ideal PRF model immune to irrelevant information should be able to generate a better revision due to more relevant information, which implies a smaller reformulation loss.
In short, the principle can be expressed as follows: 
\paragraph{Comparison Principle} Given a larger feedback set $F^{(k+1)} \supseteq F^{(k)}$, an ideal PRF model should generate a better revision $q^{(k+1)}$ whose reformulation loss is less, i.e., $\mathcal{L}_{\mathrm{rf}}(q^{(k+1)}) \le \mathcal{L}_{\mathrm{rf}}(q^{(k)})$.

The above principle describes a necessary condition for an ideal PRF model, i.e., a regular comparative relationship.
Therefore, we try to constrain this comparative relationship, which was ignored by the previous work, by means of regularization.
Instead of optimizing different revisions of the same query independently, we optimize them collectively with a comparative regularization.
First, to allow comparison between multiple revisions, at each epoch, we randomly sample $|K| > 1$ distinct PRF depths from the full set $A$ without replacement.
Then, $|K|$ revisions $\{q^{(k)}|k \in K\}$ of the same query $q$ are reformulated in parallel by the PRF model in the same batch, and their reformulation losses are calculated as $\{\mathcal{L}(q^{(k)})|k \in K\}$.
Finally, we regularize these losses to pursue the above principle during training.
Specifically, we add the following comparative regularization term to the reformulation losses,
\begin{equation}
  \mathcal{L}_{\mathrm{cr}}(q) = \frac{1}{{|K| \choose 2}}\sum_{j, k \in K \atop j < k} \max\big(0, \mathcal{L}_{\mathrm{rf}}(q^{(k)}) - \mathcal{L}_{\mathrm{rf}}(q^{(j)})\big). \label{eq:cr}
\end{equation}

As shown in Figure~\ref{fig:lol}, the regularization term $\mathcal{L}_{\mathrm{cr}}(q)$ can be viewed as a pairwise hinge~\cite{herbrich2000} ranking loss of these reformulation losses, where the ascending order of these losses is expected to coincide with the descending order of the feedback amounts they use.
Intuitively, it is reasonable if the revision $q^{(k)}$ using a larger feedback set obtains no larger reformulation loss than the revision $q^{(j)}$ using a smaller feedback set, and we should not penalize revision $q^{(k)}$.
Otherwise, we penalize $q^{(k)}$ with the loss difference $\mathcal{L}_{\mathrm{rf}}(q^{(k)}) - \mathcal{L}_{\mathrm{rf}}(q^{(j)})$ between them, while rewarding $q^{(j)}$ at the same time.
With such comparative regularization, we expect the PRF model to learn to keep the reformulation loss non-increasing with respect to the size of the feedback set by comparing the losses (effects) of different revisions and consequently learn to suppress increased irrelevant information from a larger feedback set.

In summary, the loss function of LoL consists of the conventional reformulation loss and the newly proposed comparative regularization term, formally written as follows,
\begin{equation}
  \mathcal{L}(q) = \frac{1}{|K|}\sum_{k \in K}\mathcal{L}_{\mathrm{rf}}(q^{(k)}) + \lambda\mathcal{L}_{\mathrm{cr}}(q),\label{eq:lol}
\end{equation}
where $\lambda$ is a weight that adjusts the strength of the comparative regularization. 
When we set $\lambda$ to 0 and $|K|$ to 1, Equation \eqref{eq:lol} can degenerate to Equation \eqref{eq:old}.

\subsection{A Differentiable PRF Method under LoL}

The LoL framework can be used for the training of any PRF model as long as the query reformulation loss is differentiable.
Here, we present a simple LoL implementation for both dense and sparse retrieval.

\subsubsection{Query Reformulation Loss}

The ultimate goal of PRF is to improve retrieval effectiveness. 
Generally, given a textual query $q$ and a document $d$, the similarity score between them in a single-representation retrieval model is calculated as the dot product of their vectors:
\begin{equation}
  f(q, d) = \boldsymbol{q} \cdot \boldsymbol{d},
\end{equation}
where $\boldsymbol{q}$ and $\boldsymbol{d}$ are their encoded vector representations.
In dense retrieval, $\boldsymbol{q}$ and $\boldsymbol{d}$ are dense low-dimensional vectors, while in sparse retrieval, they are sparse high-dimensional vectors whose dimensions are the size of the vocabulary.
Notably, PRF only improves the representation of the query, while the vector representation of all documents in the collection keeps invariant.

To ensure that the training objective of query reformulation is consistent with the ultimate objective of PRF, we define the reformulation loss for a revision $q^{(k)}$ derived from the original query $q$ as a ranking loss:
\begin{equation}
  \mathcal{L}_{\mathrm{rf}}(q^{(k)}) = -\log\frac{e^{f(q^{(k)}, d^+)}}{e^{f(q^{(k)}, d^+)} + \sum_{d^- \in D^-} e^{f(q^{(k)}, d^-)}},\label{eq:rf}
\end{equation}
where $d^+$ is the positive document relevant to $q$ and $q^{(k)}$, and $D^-$ is the collection of negative documents for them.
Optimizing this reformulation loss is trivial for dense retrieval, which inherently operates in the vector space.
However, since natural language queries are non-differentiable, optimizing this loss for sparse retrieval is tricky.

Considering that the query text will eventually be encoded as a vector at retrieval time, we skip the generation of the query text and directly refine the hidden representation of the query in the vector space as in dense retrieval approaches~\cite{karpukhin2020dpr, xiong2020ance}.
In other words, the vector $q^{(k)}$ refined by the PRF model $\mathrm{QR}$, hereafter we call it $\boldsymbol{q}^{(k)}$, will serve as the vector of the reformulated query in the second-pass retrieval.
In this way, we eliminate both the natural language generator that generates the textual revision and the query encoder that encodes the revised text. 
More importantly, we guarantee the differentiability of the reformulation loss in Equation \eqref{eq:rf}, which allows us to train the PRF model end-to-end.

\subsubsection{PRF Model}

In the following, we describe a simple instance of the PRF model $\mathrm{QR}$ in Equation \eqref{eq:qr}, which encodes the original query and the feedback set into a sparse or dense revision vector.

In general, the PRF model consists of an encoder, a vector projector, and a pooler.
We first leverage a BERT-style encoder to encode all tokens in the original query $q$ and the feedback set $F^{(k)}$ into contextualized embeddings:
\begin{equation}
    \boldsymbol{H} = \mathrm{BERT}(\verb|[CLS]| \circ q \circ \verb|[SEP]| \circ d_1 \circ \verb|[SEP]| \circ \ldots \circ d_k \circ \verb|[SEP]|).
\end{equation}
Then, contextualized embeddings $\boldsymbol{H}$ are projected to vectors with the same dimension as indexed document vectors: 
\begin{equation}
    \boldsymbol{V} = \mathrm{projector}(\boldsymbol{H}).
\end{equation}
Finally, all vectors are pooled into a single vector representation:
\begin{equation}
    \boldsymbol{q^{(k)}} = \mathrm{pooler}(\boldsymbol{V}).
\end{equation}

For sparse retrieval, the projector is a MLP with GeLU activation and layer normalization, initialized from a pre-trained masked language model layer.
And the pooler is composed of a max pooling operation and a L2 normalization\footnote{We experimentally find that L2 normalization helps the model train stably, and it does not change the relevance ranking of documents to a query.}.
While for dense retrieval, the projector is a linear layer, and the pooler applies a layer normalization on the first vector (\verb|[CLS]|) in the sequence, as in the previous work~\cite{xiong2020ance, yu2021anceprf}.


\section{Experimental Setup}

This section describes the datasets, evaluation metrics, baselines, and details of our implementations.

\subsection{Datasets}

Experiments are conducted on MS MARCO passage~\cite{nguyen2016} collection, which includes 8.8M English passages from web pages gathered from Bing’s results to 1M real-world queries.
We train our models with the MS MARCO Train set, which includes 530K queries with shallow annotation (\textasciitilde1.1 relevant passages per query in average).
The trained models are first evaluated on the MS MARCO Dev set containing 6980 queries to tune hyper-parameters and select model checkpoints.
We then evaluate the selected models on the MS MARCO online Eval set and three offline benchmarks (DL-HARD~\cite{mackie2021}, TREC DL 2019~\cite{craswell2020} and TREC DL 2020~\cite{craswell2021}).
MS MARCO Eval\footnote{\url{https://microsoft.github.io/MSMARCO-Passage-Ranking-Submissions/leaderboard/}}, TREC DL 2019, TREC DL 2020 and DL-HARD contain 6837, 43, 54 and 50 labeled queries, respectively.
Unlike MS MARCO, whose relevance judgments are binary, the other three benchmarks provide fine-grained annotations (relevance grades from 0 to 3) for each query.
Among them, DL-HARD~\cite{mackie2021} is a recent evaluation dataset focusing on challenging and complex queries.

\subsection{Evaluation Metrics}
The official metric of MS MARCO~\cite{nguyen2016} is MRR@10, and the main metric of TREC DL~\cite{craswell2020, craswell2021} and DL-HARD~\cite{craswell2021} is NDCG@10. 
MRR@10 is also the criterion used to select our models.
Since we focus on PRF for first-stage retrieval, we report Recall@1K for all benchmarks. 
To compute the recall metric for TREC DL and DL-HARD, we treat documents with relevance grade 1 as irrelevant following~\cite{craswell2020, craswell2021}.
To evaluate the robustness of PRF methods, we use the robustness index (RI)~\cite{collins-thompson2009}.
RI is defined as $\frac{N_{+} - N_{-}}{|Q|}$, where $|Q|$ is the total number of queries and $N_+$ and $N_-$ are the number of queries that are improved or degraded by the PRF method.
The value of RI is always in the [-1, 1] interval, and methods with higher values are more robust.
Statistically significant differences in performance are determined using the paired t-test.

\subsection{Baselines}
Since in this paper we only provide one implementation of LoL for single-representation retrieval, we do not consider baselines of re-ranking and multi-representation retrieval.

For base retrieval models without feedback, we mainly consider BM25~\cite{robertson2009bm25}, uniCOIL + docT5query~\cite{lin2021unicoil}, and ANCE~\cite{xiong2020ance}.
The first two are lexical sparse retrieval models, while ANCE is a single-representation dense retrieval model.
\begin{itemize}
  \item \textbf{BM25}~\cite{robertson2009bm25} is heuristic bag-of-words retrieval model with frequency-based signals to estimate the weights of terms in a text.
  \item \textbf{uniCOIL + docT5query}~\cite{lin2021unicoil} is a state-of-the-art trainable BERT-based term weighting model that encodes both queries and documents as 30522-dimensional sparse vectors, where each dimension is a term tokenized by WordPiece. Instead of PRF for query expansion, it expands all documents before indexing through docT5query~\cite{nogueira2019}, which leverages the powerful T5~\cite{raffel2020t5} language model to generate queries for document expansion.
  \item \textbf{ANCE}~\cite{xiong2020ance} is a popular RoBERTa-based~\cite{liu2019roberta} dense retrieval model that learns to generate a single-representation vector for each query and document via mining hard negatives from asynchronously updated document index built by the latest model checkpoint.
\end{itemize}

For the PRF models, we consider three heuristic methods (RM3, Rocchio and Average) and one supervised learning method (ANCE-PRF) based on the retrieval model described above.
\begin{itemize}
  \item \textbf{RM3}~\cite{jaleel2004rm3} is an effective relevance model for traditional sparse retrieval. 
  We apply RM3 on BM25, serving as a representative method for classic PRF.
  \item \textbf{Rocchio}~\cite{rocchio1971} and \textbf{Average} are the other two heuristic PRF methods, and have been investigated for ANCE by \cite{li2021b}.
  \item \textbf{ANCE-PRF}~\cite{yu2021anceprf} is currently the strongest PRF method for single-representation retrieval.
  Keeping the document index of ANCE unchanged, ANCE-PRF is trained end-to-end with relevance labels and learns to optimize the revised query vector by exploiting the relevant information in the feedback documents.
\end{itemize}

We also evaluate two variants of standard LoL~($\lambda > 0, |K| > 1$):
\begin{itemize}
  \item \textbf{LoL w/o Reg} ($\lambda = 0, |K| > 1$) does not impose the comparative regularization in Equation \eqref{eq:cr} but has multiple parallel revisions derived from the same query in each batch.
  \item \textbf{LoL w/o Par} ($\lambda = 0, |K| = 1$) does not revise the same original query multiple times in parallel in each batch, but unlike ANCE-PRF, it is still trained jointly for all PRF depths.
\end{itemize}

\begin{table*}[ht]
\centering
\caption{Overall retrieval results.
The best results in each group are marked in bold.
We reproduce all baseline results, except for ANCE-PRF, Rocchio, and Average, whose results are reported in previous work and not available for significance tests.
Superscript $\ast$ indicates statistically significant improvements over its base retrieval model with $p \le 0.05$.
}
\label{tab:overall}
\scalebox{0.83}{
\begin{tabular}{l|l|ccc|c|cc|cc|cc} 
\toprule
\multicolumn{2}{c|}{\textbf{Model}}                                                           & \multicolumn{3}{c|}{\textbf{MARCO Dev}} & \textbf{MARCO Eval} & \multicolumn{2}{c|}{\textbf{TREC DL 2019}} & \multicolumn{2}{c|}{\textbf{TREC DL 2020 }} & \multicolumn{2}{c}{\textbf{DL-HARD }}  \\ 
\cmidrule(lr){1-2}\cmidrule(lr){3-5}\cmidrule(lr){6-6}\cmidrule(lr){7-8}\cmidrule(lr){9-10}\cmidrule(lr){11-12}
\textbf{Retrieval}                                                           & \textbf{PRF} & NDCG@10 & MRR@10 & R@1K                  & MRR@10               & NDCG@10 & R@1K                              & NDCG@10 & R@1K                              & NDCG@10 & R@1K                         \\ 
\hline
\multirow{2}{*}{BM25~\cite{robertson2009bm25}}                                                         & -             & \textbf{23.40}   & \textbf{18.74}  & 85.73                 & \textbf{18.60}                 & 49.73   & 74.50                             & \textbf{48.76}   & 80.31                             & \textbf{28.97}   & 67.83                        \\
                                                                              & RM3~\cite{jaleel2004rm3}           & 21.35   & 16.68  & \textbf{86.86}$^\ast$               & -                    & \textbf{52.31}$^\ast$   & \textbf{77.92}$^\ast$                             & 48.08   & \textbf{82.86}$^\ast$                             & 26.63   & \textbf{69.47}$^\ast$                        \\ 
\hline
\multirow{2}{*}{\begin{tabular}[c]{@{}l@{}}uniCOIL+\\docT5query~\cite{lin2021unicoil}\end{tabular}} & -             & 41.21   & 35.13  & 95.81                 & 34.42                 & 70.09   & 82.83                             & 67.35   & 84.42                             & 35.96   & 76.85                        \\
                                                                              & LoL$^{(3)}$        & \textbf{42.02}$^\ast$   & \textbf{35.75}$^\ast$  & \textbf{96.91}$^\ast$              & \textbf{35.14}                     & \textbf{70.10}   & \textbf{83.58}                          & \textbf{69.70}   & \textbf{84.51}                             & \textbf{36.90}   & \textbf{77.67}                        \\ 
\hline
\multirow{6}{*}{ANCE~\cite{xiong2020ance}}                                                         & -             & 38.76   & 33.01  & 95.84                 & 31.70                 & 64.76   & 75.70                             & 64.58   & 77.64                             & 33.39   & 76.65                        \\
                                                                              & Average$^{(3)}$~\cite{li2021b}    & -       & -      & 94.90                 & -                    & -       & 77.39                             & -       & 79.09                             & -       & -                            \\
                                                                              & Rocchio$^{(5)}$~\cite{li2021b}    & -       & -      & 95.45                 & -                    & -       & 78.25                             & -       & 79.57                             & -       & -                            \\
                                                                              & ANCE-PRF$^{(3)}$~\cite{yu2021anceprf}   & 40.10    & 34.40   & 95.90                  & 33.00                 & 68.10    & 79.10                              & 69.50    & 81.50                              & 36.50    & 76.10                         \\
                                                                              & LoL$^{(3)}$        & 40.68$^\ast$   & 34.84$^\ast$  & 96.94$^\ast$                 & -                    & 68.42   & 80.10$^\ast$                             & 69.58$^\ast$   & 81.77$^\ast$                             & 35.61   & 79.39$^\ast$                        \\
                                                                              & LoL$^{(5)}$        & \textbf{41.01}$^\ast$   & \textbf{35.14}$^\ast$  & \textbf{97.03}$^\ast $        &     \textbf{34.17}                 & \textbf{69.58}$^\ast$   & \textbf{80.81}$^\ast$                             & \textbf{70.44}$^\ast$   & \textbf{82.77}$^\ast$                             & \textbf{37.44}$^\ast$   & \textbf{79.55}$^\ast$                        \\
\bottomrule
\end{tabular}
}
\end{table*}

\subsection{Implementation Details}
To ensure that gradients can be back-propagated during training, we perform real-time retrieval by multiplying the query matrix and the document matrix\footnote{For sparse retrieval, the document matrix is stored in sparse formats of PyTorch.}.
The query matrix consists of all revised query vectors in a batch, and the document matrix contains pre-computed vectors of all positive and mined negative documents.
These negative documents are the union of the top-ranked documents retrieved by BM25, uniCOIL + docT5query, and ANCE.
At training time, $D^{-}$ in Equation \eqref{eq:rf} contains all documents in the document matrix except the relevant documents for that query.
Since document vectors do not need to be updated, we can mine as many negative documents as possible, as long as it does not exceed the memory limit of GPUs or retrieval is too slow.

We train two PRF models using the presented LoL implementation on 4 Tesla V100 GPUs with 32GB memory for up to $\frac{12}{|K|}$ epochs\footnote{Regardless of $|K|$, all original queries are revised at most 12 times during training.}, one model for sparse retrieval with document expansion (uniCOIL + docT5query) and the other for dense retrieval (ANCE).
During training, one GPU is dedicated to retrieval, and the other three are used for the PRF model to revise query vectors.
The document matrices are converted from the pre-built inverted or dense indexes provided by pyserini\footnote{\url{https://github.com/castorini/pyserini}}, a wrapper of the Anserini IR toolkit~\cite{yang2017anserini} for Python.
We optimized the PRF models using the AdamW optimizer with the learning rate selected from $\{2 \times 10^{-5}, 1 \times 10^{-5}, 5 \times 10^{-6}\}$ for all PRF depths in $A = \{0, 1, 2, 3, 4, 5\}$\footnote{Since the maximum input length of a BERT-style PLM is 512, we consider up to 5 feedback documents in this work.}.
We set the feedback weight $\lambda$ to 1 and the number of comparative revisions $|K|$ to 2 if not specified.
For uniCOIL + docT5query, the PRF model is initialized from $\mathrm{BERT}_\mathrm{base}$, and the document matrix contains 3,738,207 documents.
We set the batch size (number of original queries) to $\frac{96}{|K|}$, which means the total number of revisions in a batch is always 96.
For ANCE, the PRF model is initialized from $\mathrm{ANCE}_\mathrm{FirstP}$\footnote{\url{https://github.com/microsoft/ANCE}}, the document matrix contains 5,284,422 documents, and the batch size is set to $\frac{108}{|K|}$.
We keep the model checkpoint with the best MRR@10 score on the MS MARCO Dev set. 
In inference, we first obtain top-ranked documents using the base retrieval model. 
Then we feed them into the trained PRF model in Equation \eqref{eq:qr} to get a revised query vector, and perform the second-pass retrieval for the final results.

For baselines BM25 and BM25 with RM3, we set $k_1$ to 0.82 and $b$ to 0.6 and reproduce them via pyserini.
For uniCOIL + docT5query or ANCE, we convert its pre-built document index to a sparse or dense matrix and reproduce its retrieval results through our matrix multiplication on GPUs.

\section{Experimental Results}

In this section, we discuss our experimental results and analysis.
We first compare LoL with typical base retrieval models and state-of-the-art PRF models;
Then, we verify the role of comparative regularization through ablation studies.
Furthermore, we investigate the robustness of LoL to PRF depths and its sensitivity to training hyper-parameters.
Finally, we visualize the impact of LoL in training through learning curves.

For simplicity, we hereafter refer to the LoL model as the PRF model optimized under the LoL framework, $\mathrm{LoL}_\mathrm{uniCOIL}$ as the LoL model for uniCOIL + docT5query, and $\mathrm{LoL}_\mathrm{ANCE}$ as the LoL model for ANCE.

\subsection{Main Results}

Table \ref{tab:overall} shows the overall retrieval results of baselines and LoL models on MARCO Dev, MARCO Eval, TREC DL 2019, TREC DL 2020 and DL-HARD.
For both sparse retrieval and dense retrieval, we report the results of LoL models at their best-performing PRF depths (numbers in superscript brackets).
For a fair comparison with ANCE-PRF$^{(3)}$, we also report the results of $\mathrm{LoL}_\mathrm{ANCE}^{(3)}$, both of which use the top 3 feedback documents.

In the first group in Table \ref{tab:overall}, we can see that RM3 improves Recall@1K of BM25 at the expense of MRR@10 and NDCG@10, which reflects the problem of query drift.

From the last two groups in Table \ref{tab:overall}, we find that all LoL models outperform their base retrieval models, i.e., uniCOIL + docT5query and ANCE, across all evaluation benchmarks and metrics.
This proves the availability of our differentiable PRF implementation of the LoL.

Compared with recent PRF baseline models for ANCE, $\mathrm{LoL}_\mathrm{ANCE}$ also achieve better retrieval performance, except for the NDCG@10 metric of $\mathrm{LoL}_\mathrm{ANCE}^{(3)}$ on the DL-HARD benchmark is lower than that of ANCE-PRF$^{(3)}$.
However, the Recall@1K of $\mathrm{LoL}_\mathrm{ANCE}^{(3)}$ on DL-HARD is improved by 4.3\% compared to ANCE-PRF$^{(3)}$, and outperforms the base ANCE without PRF.
Moreover, when five feedback documents are fed into $\mathrm{LoL}_\mathrm{ANCE}$, $\mathrm{LoL}_\mathrm{ANCE}^{(5)}$ achieves a go-ahead over ANCE-PRF$^{(3)}$ on the NDCG@10 metric.
Considering that ANCE-PRF can be regarded as a special case ($\lambda = 0$ and $|K| = |A| = 1$) under the LoL framework, the above results demonstrate the effectiveness of the Loss-over-Loss framework.

It is worth noting that even though the documents are expanded with T5-generated queries in advance, which to some extent mitigates the expression mismatch problem, $\mathrm{LoL}_\mathrm{uniCOIL}$ still improves on uniCOIL + docT5query.
This phenomenon demonstrates the powerful query reformulation capability of LoL and shows that document expansion cannot completely supplant query reformulation.

\subsection{Ablation Studies}

\begin{table}
\centering
\caption{Ablation on comparative regularization for ANCE at all PRF depths.
The best results in each group are marked in bold.
Superscripts $\dagger$, $\ddagger$ and $\S$ indicate statistically significant improvements over LoL w/o Par, LoL w/o Reg and LoL at the same PRF depth with $p \le 0.1$, respectively.
}
\label{tab:ance}
\scalebox{1}{
\begin{tabular}{l|l|ccc}
\toprule
$k$                & Method      & NDCG@10                             & MRR@10                              & R@1K           \\
\midrule
-                  & ANCE        & 38.76                               & 33.01                               & 95.84          \\
\hline
\multirow{4}{*}{0} & ANCE-PRF    & 36.40                               & 30.70                               & 94.30          \\
\cdashline{2-5}
                   & LoL w/o Par & 38.21                               & 32.47                               & 96.11          \\
                   & LoL w/o Reg & 38.25                               & 32.49                               & 96.10          \\
                   & LoL         & \textbf{38.68}$^{\dagger\ddagger}$  & \textbf{32.85}$^{\dagger\ddagger}$  & \textbf{96.22} \\
\hline
\multirow{4}{*}{1} & ANCE-PRF    & 39.30                               & 33.40                               & 96.30          \\
\cdashline{2-5}
                   & LoL w/o Par & 39.59                               & 33.77                               & 96.67          \\
                   & LoL w/o Reg & 39.56                               & 33.73                               & \textbf{96.82} \\
                   & LoL         & \textbf{39.82}$^{\dagger\ddagger}$  & \textbf{33.98}$^{\dagger\ddagger}$  & 96.77          \\
\hline
\multirow{4}{*}{2} & ANCE-PRF    & 40.10                               & 34.30                               & 96.20          \\
\cdashline{2-5}
                   & LoL w/o Par & 40.27                               & 34.46                               & 96.83          \\
                   & LoL w/o Reg & 40.16                               & 34.33                               & 96.79          \\
                   & LoL         & \textbf{40.39}$^{\ddagger}$         & \textbf{34.51}                      & \textbf{96.86} \\
\hline
\multirow{4}{*}{3} & ANCE-PRF    & 40.10                               & 34.40                               & 95.90          \\
\cdashline{2-5}
                   & LoL w/o Par & 40.58                               & 34.71                               & 96.94          \\
                   & LoL w/o Reg & 40.40                               & 34.62                               & 96.83          \\
                   & LoL         & \textbf{40.68}$^{\ddagger}$         & \textbf{34.84}$^{\ddagger}$         & \textbf{96.94} \\
\hline
\multirow{4}{*}{4} & ANCE-PRF    & 40.30                               & 34.60                               & 96.10          \\
\cdashline{2-5}
                   & LoL w/o Par & 40.59                               & 34.72                               & 96.93          \\
                   & LoL w/o Reg & 40.53                               & 34.66                               & 96.90          \\
                   & LoL         & \textbf{40.83}$^{\dagger\ddagger}$  & \textbf{34.95}$^{\ddagger}$         & \textbf{97.01} \\
\hline
\multirow{4}{*}{5} & ANCE-PRF    & 40.00                               & 34.40                               & 96.00          \\
\cdashline{2-5}
                   & LoL w/o Par & 40.72                               & 34.84                               & 96.96          \\
                   & LoL w/o Reg & 40.77                               & 34.85                               & 96.93          \\
                   & LoL         & \textbf{41.01}$^{\dagger\ddagger}$  & \textbf{35.14}$^{\dagger\ddagger}$  & \textbf{97.03} \\
\bottomrule
\end{tabular}
}
\end{table}

\begin{table}
\centering
\caption{Ablation on comparative regularization for uniCOIL + docT5query at all PRF depths.
The best results in each group are marked in bold.
Superscripts $\dagger$, $\ddagger$ and $\S$ indicate statistically significant improvements over LoL w/o Par, LoL w/o Reg and LoL at the same PRF depth with $p \le 0.1$, respectively.
}
\label{tab:uncoil}
\scalebox{1}{
\begin{tabular}{l|l|ccc} 
\toprule
$k$                  & Method             & NDCG@10 & MRR@10 & R@1K   \\
\midrule
-                  & uniCOIL+docT5query & 41.21   & 35.13  & 95.81  \\
\hline
\multirow{3}{*}{0} & LoL w/o Par        & 41.31                               & 35.03                               & 96.69  \\
                   & LoL w/o Reg        & 41.13                               & 34.92                               & 96.79  \\
                   & LoL                & \textbf{41.36}$^{\ddagger}$         & \textbf{35.08}$^{\ddagger}$         & \textbf{96.80}$^{\dagger}$  \\
\hline
\multirow{3}{*}{1} & LoL w/o Par        & 41.73                               & 35.46                               & 96.79   \\
                   & LoL w/o Reg        & 41.73                               & 35.50                               & 96.95   \\
                   & LoL                & \textbf{41.86}                      & \textbf{35.62}                      & \textbf{96.98}$^{\dagger}$  \\
\hline
\multirow{3}{*}{2} & LoL w/o Par        & 41.83                               & 35.56                               & 96.82   \\
                   & LoL w/o Reg        & 41.81                               & 35.59                               & \textbf{97.06}  \\
                   & LoL                & \textbf{41.94}$^{\ddagger}$         & \textbf{35.68}                      & 97.01$^{\dagger}$  \\
\hline
\multirow{3}{*}{3} & LoL w/o Par        & 41.76                               & 35.48                               & 96.85   \\
                   & LoL w/o Reg        & 41.75                               & 35.51                               & \textbf{97.03}$^{\S}$  \\
                   & LoL                & \textbf{42.02}$^{\dagger\ddagger}$  & \textbf{35.75}$^{\dagger\ddagger}$  & 96.91   \\
\hline
\multirow{3}{*}{4} & LoL w/o Par        & 41.61                               & 35.28                               & 96.85   \\
                   & LoL w/o Reg        & 41.74                               & 35.43                               & \textbf{97.03}  \\
                   & LoL                & \textbf{41.94}$^{\dagger\ddagger}$  & \textbf{35.67}$^{\dagger\ddagger}$  & 96.96$^{\dagger}$   \\
\hline
\multirow{3}{*}{5} & LoL w/o Par        & 41.68                               & 35.37                               & 96.87  \\
                   & LoL w/o Reg        & 41.74                               & 35.44                               & \textbf{97.04}  \\
                   & LoL                & \textbf{41.94}$^{\dagger\ddagger}$  & \textbf{35.67}$^{\dagger\ddagger}$  & 96.89  \\
\bottomrule
\end{tabular}
}
\end{table}

In this part, we conduct ablation studies on MARCO Dev for both sparse and dense retrieval to further explore the roles of comparative regularization and multiple parallel revisions in LoL.

A standard LoL ($\lambda = 1,|K| = 2$) and two LoL variants, i.e., LoL w/o Reg ($\lambda = 0,|K| = 2$) and LoL w/o Par ($\lambda = 0,|K| = 1$), are measured at all PRF depths in $A$.
We compare the evaluation results of the standard LoL and LoL w/o Reg to show the role of the comparative regularization in Equation \eqref{eq:cr}.
We further introduce LoL w/o Par to eliminate the effect of parallel revision multiple times in one batch.
For dense retrieval, we also compare LoL models to ANCE-PRF models, which are equivalent to LoL w/o Par trained separately at each PRF depth.
Note that we use different checkpoints for the model of the same type at different PRF depths, which are selected for each PRF depth separately.

As shown in Table \ref{tab:ance}, at each PRF depth, the standard LoL outperforms its two variants and ANCE-PRF in all metrics, with the one exception of recall@1K at $k = 1$, where LoL w/o Reg is slightly better than LoL.
The conclusions of the sparse search in Table \ref{tab:uncoil} are similar, although there are four slight drops in recall compared to LoL w/o Reg.
We speculate this may be because the ranking loss function in Equation \eqref{eq:rf} is closer to the shallower metrics like NDCG@10 and MRR@10.
And the comparative regularization further increases LoL models' attention to these shallow ranking metrics.
Therefore, it is sufficient to show the effectiveness of the comparative regularization.
In addition, we find that LoL w/o Reg and LoL w/o Par are generally competitive with each other, which indicates the impact of parallel multiple revisions is not significant and highlights the role of comparative regularization.

Moreover, as shown in Table \ref{tab:ance}, LoL w/o Par also outperforms the ANCE-PRF across the board, especially the Recall@1K metric.
We believe this may be attributed to joint training and the computation of reformulation loss on the entire (mined) document matrix.

\subsection{Robustness to PRF Depth}

\begin{figure}[ht]
    \centering
    \begin{subfigure}[b]{0.45\textwidth}
    \centering
    \includegraphics[width=\textwidth]{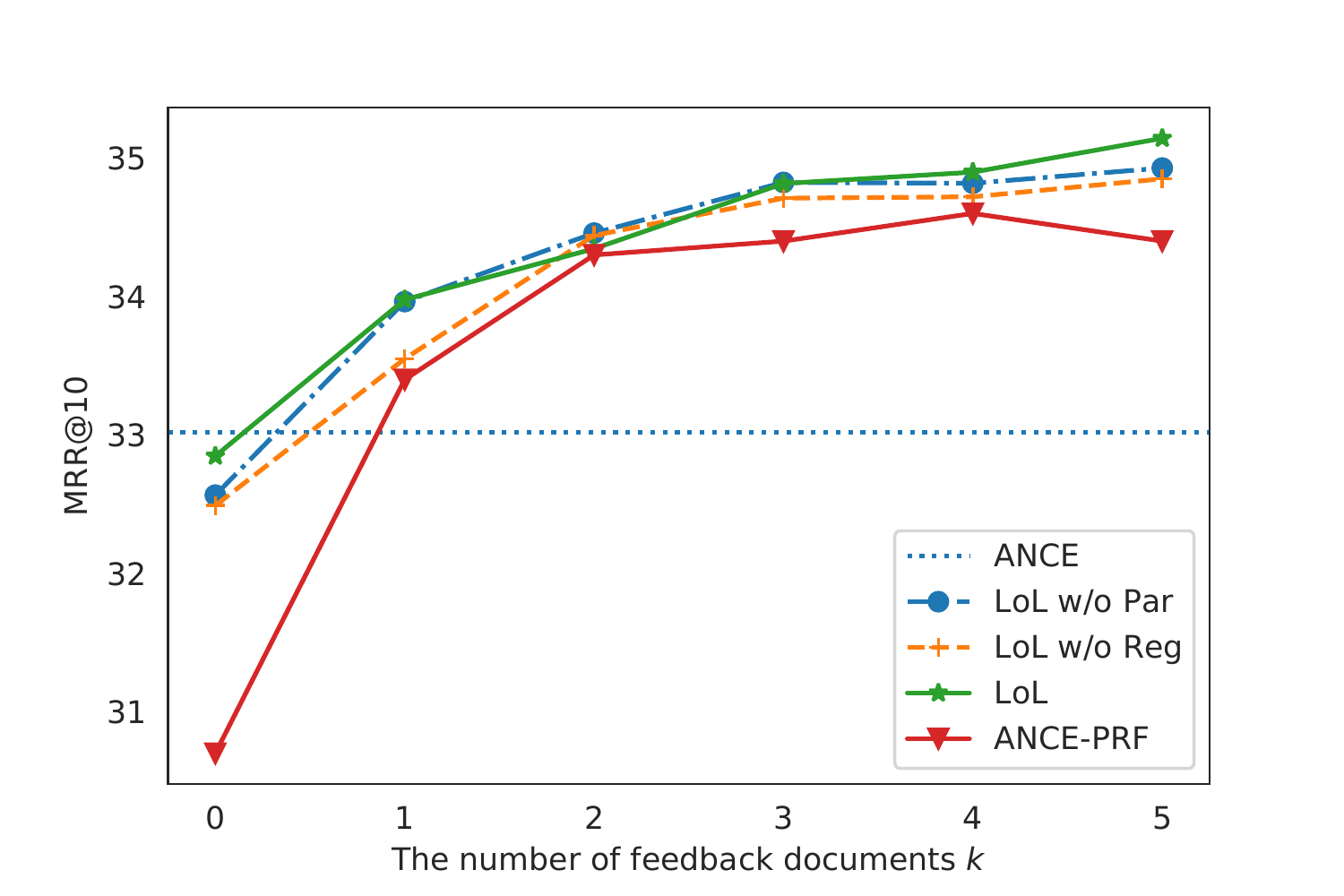}
    \caption{ANCE}
    \label{fig:ance_depths}
    \end{subfigure} 
    \begin{subfigure}[b]{0.45\textwidth}
    \centering
    \includegraphics[width=\textwidth]{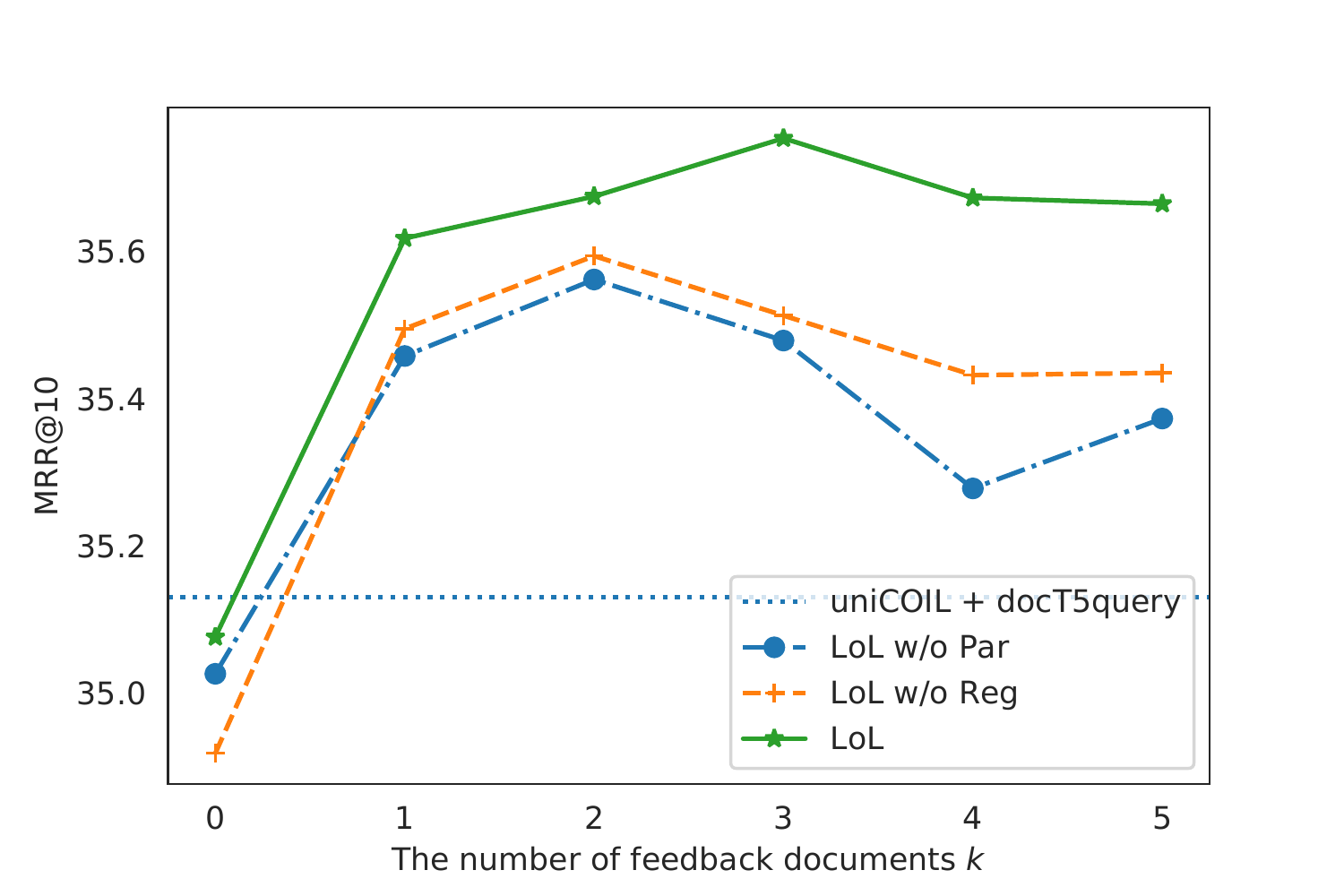}
    \caption{uniCOIL + docT5query}
    \label{fig:unicoil_depths}
    \end{subfigure} 
    \caption{The MRR@10 curve of the same PRF model using different numbers of documents. The horizontal dotted line represents the MRR@10 value of the base retrieval model. 
    }
    \label{fig:prf_depths}
\end{figure}

At the beginning of the design, we expect LoL to alleviate query drift, i.e., make the model more robust to the increasing number of feedback documents.
In this part, we verify this expectation.

Figure~\ref{fig:prf_depths} shows the performance of the best checkpoint for multiple PRF models at all PRF depths.
Different from using different model checkpoints at different PRF depths in Table~\ref{tab:ance} and Table~\ref{tab:uncoil}, each curve of LoL in Figure~\ref{fig:prf_depths} is drawn from the performance of the same model checkpoint.
Therefore, the MRR@10 values in Table~\ref{tab:ance} and Table~\ref{tab:uncoil} can be viewed as the upper bound of the values in Figure~\ref{fig:ance_depths} and Figure~\ref{fig:unicoil_depths}, respectively.
As we can see in Figure~\ref{fig:ance_depths} and Table~\ref{tab:ance}, only LoL and LoL w/o Reg are monotonically increasing with respect to the number of documents.
ANCE-PRF reaches peak performance at PRF depth 4 and then suffers performance degradation, and LoL w/o Par. encounters a performance dip when the number of feedback documents increased from 3 to 4.
As for PRF models applied in sparse retrieval in Figure~\ref{fig:unicoil_depths} and Table~\ref{tab:uncoil}, LoL w/o Par. and LoL w/o Reg reach peak performance at PRF depth 2, while LoL continues to grow until the PRF depth approaches 4.

To quantify the robustness of LoL, we report the robustness indices (RI) of $\mathrm{LoL}_\mathrm{ANCE}^{(k)}$ with respect to ANCE and $\mathrm{LoL}_\mathrm{ANCE}^{(k-1)}$ in Table~\ref{tab:ri-base} and Table~\ref{tab:ri-less}, respectively.
From Table~\ref{tab:ri-base}, we can find that $\mathrm{LoL}_\mathrm{ANCE}$ reformulates more revisions that are better than original queries compared to its variant baselines at all PRF depths.
Similarly, as shown in Table~\ref{tab:ri-less}, when the number of feedback documents increases from $k-1$ to $k$, compared to LoL w/o Par and LoL w/o Reg, LoL can more robustly revise better queries than before.

From these observations, we may draw two conclusions.
(1) Compared to these baselines, LoL is more robust to PRF depths. 
That is, as the number of feedback documents increases, LoL-reformulated queries have less drift and are less prone to performance degradation.
(2) LoL for dense retrieval is more robust than LoL for sparse retrieval.
We conjecture that this is because dense query vectors are more fine-grained and are more likely to prevent the introduction of irrelevant information, while sparse query vectors are term-grained and may introduce relevant polysemous terms when reformulating the query, which in turn leads to query drift.

\begin{table}[ht]
\centering
\caption{RI of $\mathrm{LoL}_\mathrm{ANCE}^{(k)}$ with respect to ANCE on MARCO Dev at all PRF depths.}
\label{tab:ri-base}
\begin{tabular}{lcccccc} 
\toprule
\multicolumn{1}{c}{$k$} & 0             & 1             & 2             & 3             & 4             & 5              \\ 
\midrule
LoL w/o Par             & 0.34          & 0.42          & 0.41          & 0.42          & 0.41          & 0.41           \\
LoL w/o Reg             & 0.32          & 0.40          & 0.41          & 0.41          & 0.41          & 0.40           \\
LoL                     & \textbf{0.36} & \textbf{0.43} & \textbf{0.43} & \textbf{0.44} & \textbf{0.44} & \textbf{0.44}  \\
\bottomrule
\end{tabular}
\end{table}

\begin{table}[ht]
\centering
\caption{RI of $\mathrm{LoL}_\mathrm{ANCE}^{(k)}$ with respect to $\mathrm{LoL}_\mathrm{ANCE}^{(k-1)}$ on MARCO Dev at all PRF depths.}
\label{tab:ri-less}
\begin{tabular}{lccccc} 
\toprule
\multicolumn{1}{c}{$k$} & 1             & 2             & 3             & 4             & 5              \\ 
\midrule
LoL w/o Par             & 0.51          & 0.54          & 0.58          & 0.58          & 0.61           \\
LoL w/o Reg             & 0.52          & 0.53          & 0.58          & 0.59          & 0.61           \\
LoL                     & \textbf{0.54} & \textbf{0.56} & \textbf{0.63} & \textbf{0.63} & \textbf{0.66}  \\
\bottomrule
\end{tabular}
\end{table}

\subsection{Sensitivity to Training Hyper-parameters}

To capture the sensitivity of LoL to the number of comparative revisions $|K|$ and the regularization weight $\lambda$, we evaluate multiple $\mathrm{LoL}_\mathrm{ANCE}$ models trained with different $|K|$ and $\lambda$ on MARCO Dev set.
As shown in Table~\ref{tab:hp}, all $\mathrm{LoL}_\mathrm{ANCE}$ models trained with $|K| > 1$ perform better than that with $|K| = 1$, which indicates that is not sensitive to $|K| > 1$.
Comparing the last two rows that share the regularization weight $\lambda = 0.5$, we can find that the smaller $|K|$ seems to be better trained than the larger $|K|$.
We speculate that this may be because larger $|K|$ leads to smaller training batch size under the GPU memory limitation.
Using the default setting of $|K| = 2$, although the variance of the values in rows 2 to 4 is not large, setting $\lambda$ to 1 performs best at most PRF depths.

\begin{table}
\centering
\caption{MRR@10 of $\mathrm{LoL}_\mathrm{ANCE}$ with different training hyper-parameters at all PRF depths on MARCO Dev.}
\label{tab:hp}
\begin{tabular}{llccccc} 
\toprule
\multirow{2}{*}{$|K|$} & \multirow{2}{*}{$\lambda$} & \multicolumn{5}{c}{$k$}                                                                                                                      \\ 
\cmidrule(r){3-7}
                                       &                         & 1                         & 2                         & 3                         & 4                         & 5                          \\ 
\midrule
1                                      & 0                       & 33.74 & 33.38 & 34.57 & 34.51 & 34.59  \\ 
\hline
2                                      & 1.5                     & 34.02                     & 34.58                     & 34.75                     & 34.89                     & 35.08                      \\
2                                      & 1                       & 33.98                     & 34.51                     & 34.84                     & 34.95                     & 35.14                      \\
2                                      & 0.5                     & 33.99                     & 34.59                     & 34.89                     & 34.98                     & 35.10                      \\
3                                      & 0.5                     & 33.97                     & 34.61                     & 34.77                     & 34.79                     & 35.05                      \\
\bottomrule
\end{tabular}
\end{table}

\subsection{Analysis of Loss Curves}

\begin{figure*}[htbp]
    \centering
    \begin{subfigure}[b]{0.33\textwidth}
    \centering
    \includegraphics[width=\textwidth]{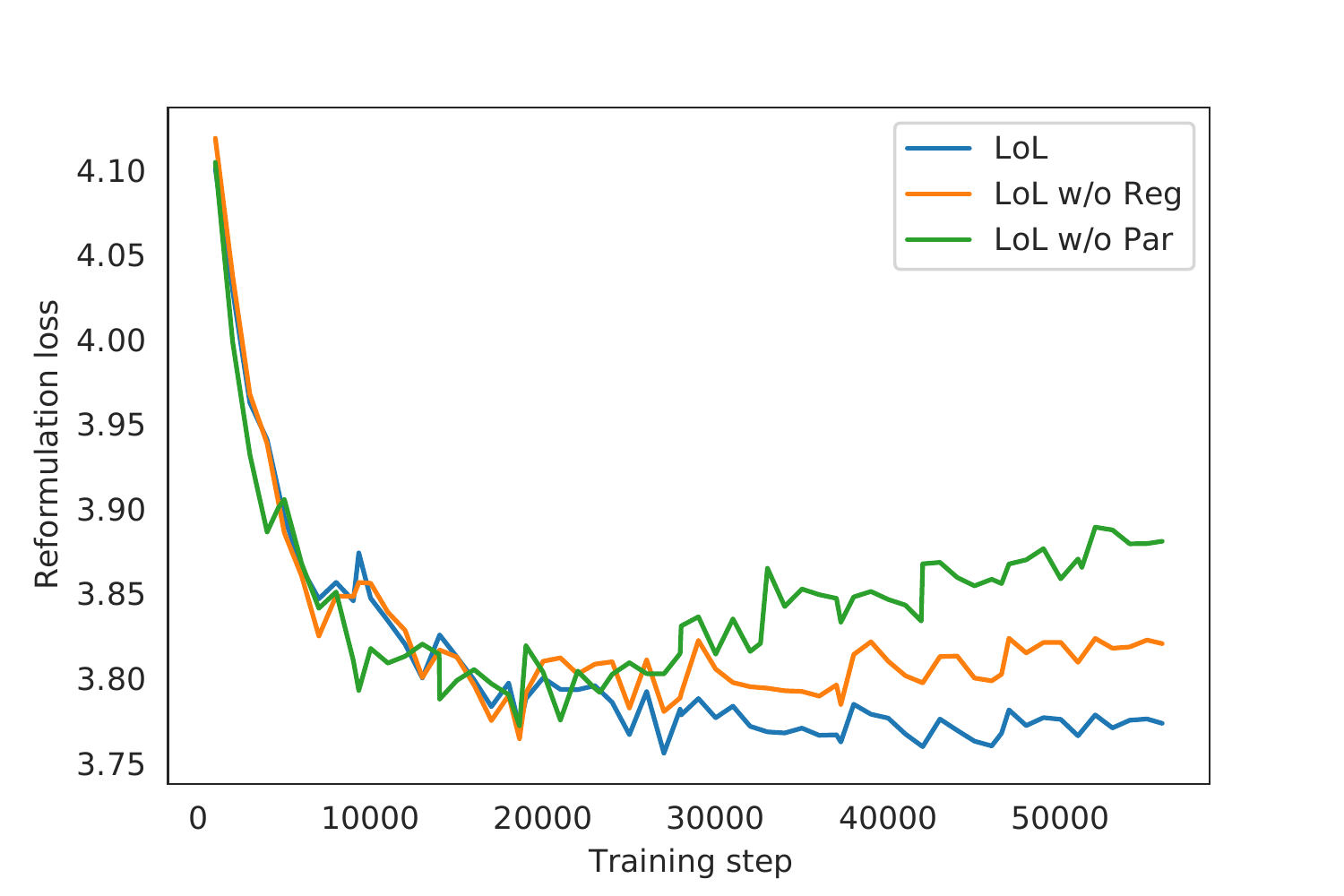}
    \caption{Reformulation loss on Dev}
    \label{fig:rf_dev}
    \end{subfigure}
    \begin{subfigure}[b]{0.33\textwidth}
    \centering
    \includegraphics[width=\textwidth]{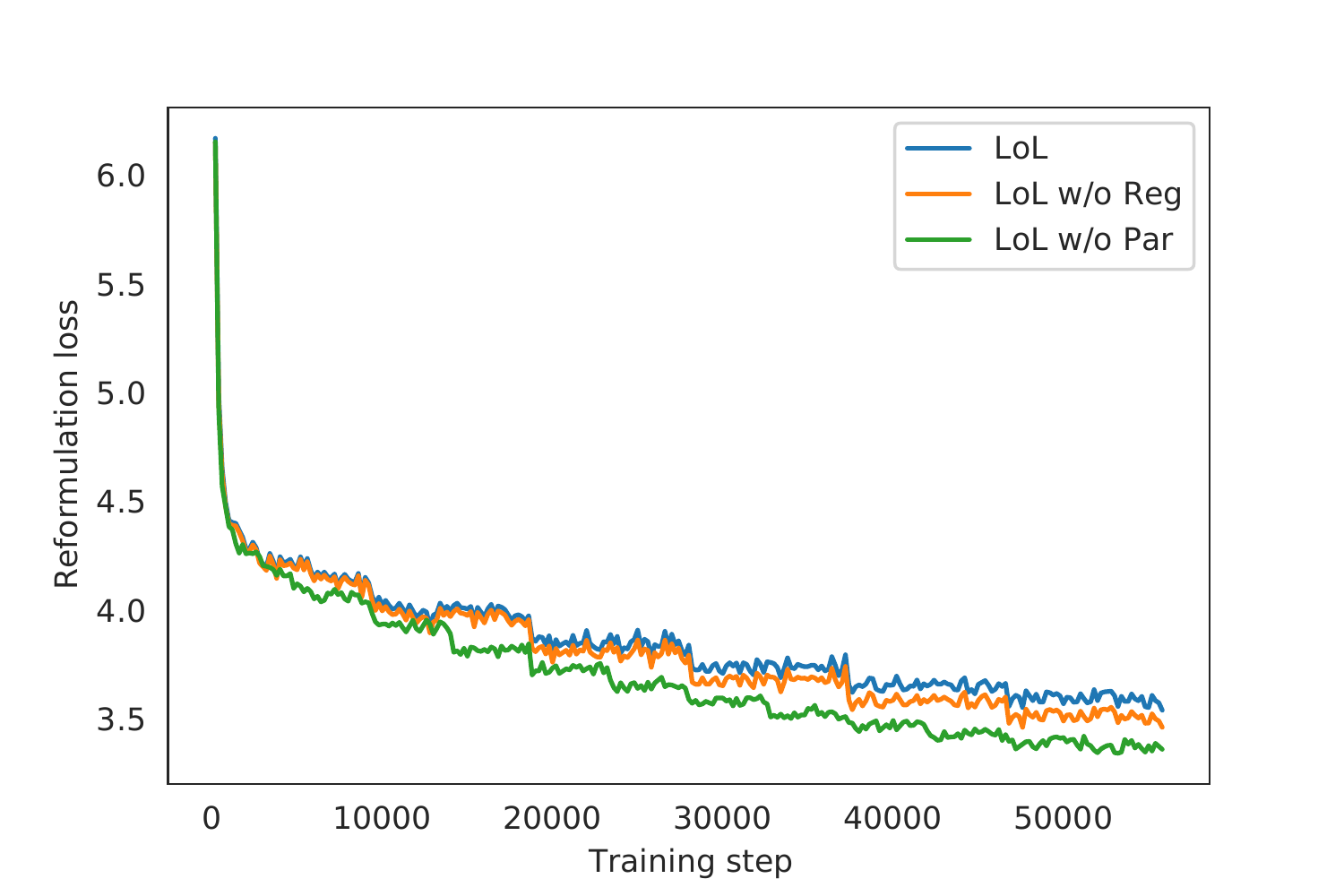}
    \caption{Reformulation loss on Train}
    \label{fig:rf_train}
    \end{subfigure}
    \begin{subfigure}[b]{0.33\textwidth}
    \centering
    \includegraphics[width=\textwidth]{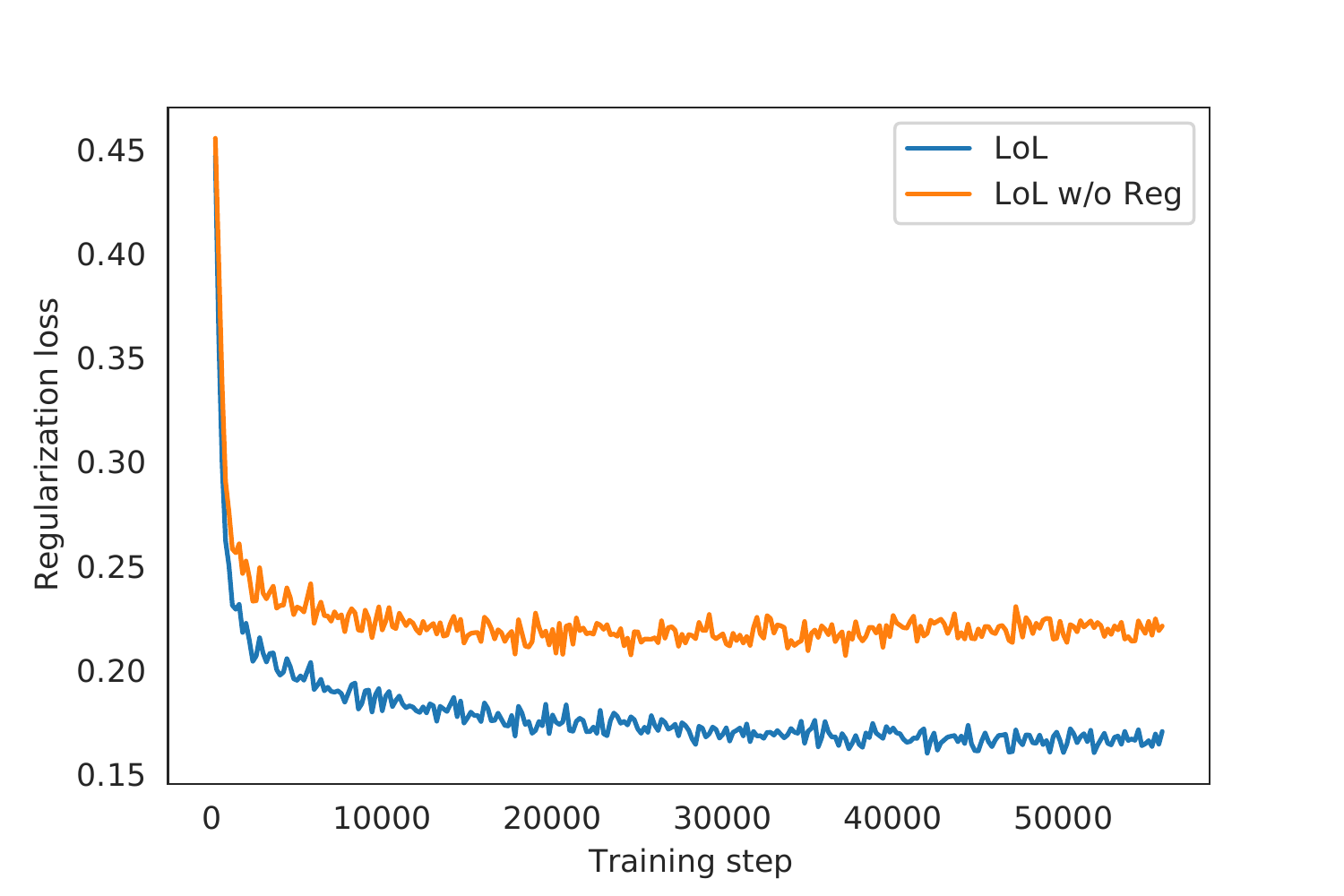}
    \caption{Regularization loss on Train}
    \label{fig:cr_train}
    \end{subfigure}
    \caption{The loss curves of standard LoL, LoL w/o Reg and LoL w/o Par for ANCE on MS MARCO.
    }
    \label{fig:losses}
\end{figure*}

To visualize the impact of LoL in training, we show the loss curves of $\mathrm{LoL}_\mathrm{ANCE}$ on the MARCO Train and Dev sets in Figure~\ref{fig:losses}.
Figure~\ref{fig:rf_dev} and \ref{fig:rf_train} plot the query reformulation losses in Equation~\eqref{eq:rf} on Train and Dev sets at training time.
We can find that the training reformulation loss of LoL w/o Par drops the fastest and lowest, followed by LoL w/o Reg, and LoL the slowest. 
And their performance on Dev set is just the opposite, where the evaluation reformulation loss of LoL w/o Par and LoL w/o Reg starts to increase successively after reaching the lowest point, while the evaluation loss of LoL rises slightly at the end.
This indicates that both comparative regularization and multiple parallel revisions have the effect of mitigating overfitting, and comparative regularization has more impact.
Figure~\ref{fig:cr_train} shows the comparative regularization terms of LoL and LoL w/o Reg.
They both revise a query multiple times in parallel.
Without regularizing the reformulation losses of these parallel revisions, the regularization loss of LoL w/o Reg also drops but not to a level as low as LoL.
This implies that a PRF model cannot naturally learn to guarantee the normal comparison relationship among multiple revisions without explicitly imposing regularization.
Even with regularization imposed, we can see that guaranteeing this normal comparison relationship is not easy for the model, because this regularization loss drops slowly in the middle and late stages of training.

\section{Discussion}
Further deriving the final loss in Equation~\eqref{eq:lol}, we can find that LoL can be viewed as re-weighting multiple reformulation losses of the same query.
For simplicity, we denote $\mathcal{L}_{\mathrm{rf}}(q^{(k)}))$ as $L^{k}$.
Speicially, the loss can be rewrited as follow:
\begin{displaymath}
\begin{aligned}
\mathcal{L}(q) &= \frac{1}{|K|} \sum_{k \in K}L^{k} + \lambda\mathcal{L}_{\mathrm{cr}}(q) \\
&= \frac{1}{|K|} \left[ \sum_{k \in K} L^k + \frac{2\lambda}{|K| - 1} \sum_{j, k \in K \atop j < k} \max(0, L^k - L^j) \right] \\
&= \frac{1}{|K|} \sum_{k} \left[ L^k + \frac{2\lambda}{|K| - 1} \sum_{j < k} \max(0, L^k - L^j) \right] \\
&= \frac{1}{|K|} \sum_{k} \left[ \left( 1 + 2\lambda\frac{\sum_{j<k}\mathds{1}(L^k > L^j)}{|K| - 1} \right)L^k - 2\lambda\frac{\sum_{i<k}\mathds{1}(L^k>L^i)}{|K|-1}L^i \right] \\
&= \frac{1}{|K|} \sum_{k} \left[ 1 + \frac{2\lambda}{{|K| - 1}} \left( \sum_{j<k}\mathds{1}(L^k>L^j) -\sum_{i>k}\mathds{1}(L^k<L^i) \right) \right] L^k \\
&= \frac{1}{|K|} \sum_{k} \left[ 1 + 2\lambda\frac{\sum_{j \neq k}\mathrm{CMP}(k, j, L^k, L^j)}{|K| - 1} \right] L^k,
\end{aligned}
\end{displaymath}
where $\mathds{1}(\cdot)$ is a indicator function and $\mathrm{CMP}$ is a function to compare the sizes of the feedback sets and evaluation losses of two revisions derived from the same query.
Formally, the $\mathrm{CMP}$ function is defined as:
\begin{displaymath}
\mathrm{CMP}(k, j, L^k, L^j) = 
\begin{cases} 
1,  & \text{if }j < k\text{ and }L^j < L^k \\
-1, & \text{if }j > k\text{ and }L^j > L^k \\
0, & \text{otherwise}.
\end{cases}
\end{displaymath}

From this re-weighting perspective, given the size of the PRF depth set $|K|$, the training complexity of LoL is the same as LoL w/o Reg ($\lambda = 0$), and the additional comparison overhead is a small constant and negligible.
Besides, since LoL is just an optimization framework, PRF models trained under LoL do not have any increase in computational cost at inference time.

Essentially, comparative regularization aims to guarantee the normal order of a set of objects. 
This normal order is usually supposed to be maintained, i.e. unsupervised, but ignored by the model. 
Therefore, from this perspective, LoL can be seen as an unsupervised application of leaning-to-rank. 
As such, one future direction is to explore the application of other leaning-to-rank losses here.
Furthermore, these objects should be able to be mapped to differentiable values, such as evaluation metrics or losses. 
Therefore, future work can also replace the mapping function (reformulation loss) in our method.
Moreover, if there are similar neglected normal orders in other tasks, then the comparative regularization may also be used for other tasks.

\section{Conclusion}
In this paper, we find that the query drift problem in pseudo-relevance feedback is mainly caused by irrelevant information when more pseudo-relevant documents are involved as feedback information to reformulate the query.
Ideally, a good pseudo-relevance feedback model should have the ability to use more feedback documents that contain irrelevant information. That is, the more pseudo-relevant documents provided, the better quality of the reformulated query.
Armed with this intuition, we design a novel comparative regularization loss based on multiple query reformulation losses to ensure that more feedback documents lead to smaller query reformulation losses.
The proposed comparative regularization loss over query reformulation losses (LoL) framework can be used in any pseudo-relevance feedback model with any retrieval framework, e.g., sparse retrieval or dense retrieval.
Experiments on publicly large-scale dataset MS MARCO and its variant evaluation sets demonstrate that our plug-and-play regularization can bring improvements compared to the baseline methods.

\begin{acks}
This work was supported by the National Natural Science Foundation of China (NSFC) under Grants No.61906180, U21B2046 and Liang Pang, Huawei Shen, Yanyan Lan are also supported by Beijing Academy of Artificial Intelligence (BAAI).
\end{acks}

\bibliographystyle{ACM-Reference-Format}
\balance
\bibliography{main}


\end{document}